\documentclass[12pt]{article}
\RequirePackage[T1]{fontenc}        
\RequirePackage[latin1]{inputenc}   
\RequirePackage{amsmath,amsfonts,graphicx}   
\usepackage{algorithm,algorithmic}

\usepackage{graphicx}
\usepackage{epsfig}
\usepackage{epsf}
\usepackage{amsmath,amssymb} 
\usepackage{rotating}

\newtheorem{thm}{Theorem}[section]

\newtheorem{defi}{Definition}[section]

\newtheorem{prop}{Proposition}[section]

\usepackage{amssymb}

\begin{document}



\title{Estimation of Gaussian mixtures in small sample studies using $l_1$ penalization }


\author{St\'ephane Chr\'etien
\thanks{Mathematics Department, Universit\'e de Franche Comt\'e and UMR CNRS-6623, 16 route de Gray, 25030 Besan{\c c}on, France. }}
\maketitle

\begin{abstract}
Many experiments in medicine and ecology can be conveniently modeled by finite Gaussian mixtures but face the problem of dealing with 
small data sets. We propose a robust version of the estimator based on self-regression and sparsity promoting penalization in order to 
estimate the components of Gaussian mixtures in such contexts. A space alternating version of the penalized EM algorithm is obtained and
we prove that its cluster points satisfy the Karush-Kuhn-Tucker conditions. Monte Carlo experiments are presented in order to compare the 
results obtained by our method and by standard maximum likelihood estimation. In particular, our estimator is seen to perform better than the maximum likelihood estimator.  
\end{abstract}

{\bf Keywords}:
finite Gaussian mixtures, maximum likelihood estimation, Kullback Proximal Point algorithms, EM algorithm, $l_1$ penalization, LASSO, sparsity, regression mixtures, 
model based clustering




\section{Introduction}
Finite Gaussian mixture models are widely used in a great number of application fields as a means to perform model based classification. From pattern recognition to biology, from quality control to finance, many examples have shown the pertinence of the Gaussian mixture model approach. The book \cite{McLachlan:Book00} is the most comprehensive reference for finite non necessarily Gaussian mixture models with many application examples. In Gaussian mixture models, the data $Y_1,\ldots,Y_n$ are assumed i.i.d. and to be drawn from  the density 
\begin{equation}
\sum_{k=1}^K p^*_k f^{(d)}(y;\mu_k,\Sigma_k)
\end{equation}
where 
\begin{eqnarray}
f^{(d)}(y;\mu,\Sigma) & = & 
\frac1{\sqrt{(2\pi)^d {\rm det} (\Sigma^*)\big)}} \exp \Big( -\frac12 (y-\mu^*)^t{\Sigma^*}^{-1}(y-\mu^*) \Big)
\end{eqnarray}
and where the vector $\theta^*=(p_1^*,\ldots,p_K^*,\mu_1^*,\ldots,\mu_K^*,\Sigma_1^*,\ldots,\Sigma_k^*)$ is an unknown multidimensional parameter. To this model, 
we traditionally associate an extended model using the notion of complete data. In mixture models, the complete data are independent and 
identically distributed couples of the form $(Y_i,Z_i)$ where $Z_i$ is a multinomial random variable taking values in $\{1,\ldots,K\}$ with $P(Z_i=k)=p^*_k$ 
and which represents the index of the mixture component from which observation $i$ was drawn. We assume that conditionally 
on the event $Z_i=k$, $Y_i$ has density $\frac1{\sqrt{(2\pi)^d {\rm det} (\Sigma_k^*)}} \exp \Big( -\frac12 (y-\mu_k^*)^t{\Sigma_k^*}^{-1}(y-\mu_k^*) \Big)$. The variables $Z_1,\ldots,Z_n$ being unobserved, they are usually called latent variables. 

The standard approach for estimating $\theta^*$ is the maximum likelihood methodology which consists of finding $\hat{\theta}$ which maximizes the log-likelihood function 
\begin{equation}
\label{lkhd}
l(\theta)=\sum_{i=1}^n \log \Big( \sum_{k=1}^K p_k f^{(d)}(y;\mu_k,\Sigma_k) \Big)   
\end{equation}
over the set 
\begin{eqnarray*}
\Theta & = & \Big\{(p_1,\ldots,p_K,\mu_1,\ldots,\mu_K,\Sigma_1,\ldots,\Sigma_K) \mid p_k \in \mathbb R_+, \: 
\mu_k\in \mathbb R^d,\: \Sigma_k \in \mathbb S_d^+, \\
 & & \hspace{3cm} \textrm{ and }\sum_{k=1}^K p_k=1 \Big\}
\end{eqnarray*} 
where $\mathbb S_d^+$ denotes the set 
of all symmetric positive semidefinite matrices and $\mathbb R_+$ is the set of nonnegative real numbers. 

Interestingly enough, the supremum of the log-likelihood function over $\Theta$ is equal to $+\infty$ and is obtained for 
singular covariance matrices. A study of the one dimensional case was made in \cite{Biernacki:SPL03}. 
However, many researchers and practitioners have noticed that some local maximizer of the log-likelihood function is in fact consistent in practice. 
From the numerical viewpoint, local maximizers of the log-likelihood function are usually obtained using the 
Expectation-Maximization (EM) algorithm 
of Dempster Laird and Rudin \cite{Dempster:JRSS77}. This algorithm is a nice procedure with closed form expression of each iteration in the Gaussian 
mixture case. The EM algorithm for mixture models is available in the MIXMOD package \cite{Biernacki:CSDA06} for instance.

Beside the question of finding the right local optimizer of the likelihood function, one of the main problems about estimating $\theta^*$ is the one of having a sufficiently large sample size. 
Instances where the sample size is large enough can be found in a number of applications such as pattern recognition or 
financial time series analysis. On the other hand, in many other fields, e.g. ecology, 
the sample size may be very small in situations where finite mixture models are suspected to be very pertinent due to the biological context. The goal of this paper is to remedy this problem by proposing a new methodology for Gaussian mixture 
model estimation in the case where the sample size is extremely small. Our approach aims at providing 
a certain amount of robustness. In the same spirit as for the median in the one dimensional case, the main idea is to express the 
estimators of the $\mu_k$'s as a combination of a small number of data in the middle of each cluster. 
This is simply done by restricting the search to the data's span, i.e. to obtain the $\mu_k$'s as a 
regression with covariates the data themselves and to 
impose an additional sparsity constraint on the regression vectors. In order to simplify the analysis, we 
will assume the covariance matrices to be of the form $\sigma_k^2 I$, where $I$ denotes the identity matrix. 
The  $\sigma_k$'s and the $p_k$'s can also be estimated using for 
instance a maximum likelihood approach conditioned on the estimated value of the $\mu_k$'s. 

The whole procedure is formally equivalent to joint variable selection and estimation in a mixture of regression model. 
Variable selection and estimation are performed using $l_1$-penalized EM steps 
which reduce the complexity of the regression model just as for the LASSO \cite{Tibshirani:JRSS96}. 
Encouraging simulations results show that the proposed approach correctly estimates the 
class of 8 over 10 points on average for a mixture of 3 Gaussians in dimension two. Monte Carlo experiments 
are performed for samples sizes of 10 points and dimension growing up to to 50 showing a good behavior of the 
method which outperforms the standard maximum likelihood estimator.

\section{Presentation of the method}

\subsection{Recalls on regression mixtures}
The Gaussian regression mixture assumes that the observations are couples of the form $(Y,X)$ where 
$Y$ takes values in $\mathbb R^d$, $X$ takes values in $\mathbb R^p$, and conditionally on $X$, the 
random variable $Y$ follows the mixture density 
\begin{equation}
f_{Y\mid X}(y) =\sum_{k=1}^K p_k \ f^{(d)}(y;M_kX,\Sigma_k),
\end{equation} 
where $M$ is a matrix in $\mathbb R^{d\times p}$. 
 
Such mixture models are frequent in econometrics and chemometrics as described in the introduction of \cite{Hurn:JCGS03}. 
Estimation in these models can be performed using likelihood maximization as in \cite{McLachlan:Book00} using the 
EM algorithm or a Bayesian methodology as studied in \cite{Hurn:JCGS03} using Monte Carlo Markov Chain techniques. 

One way to perform model selection in such a model is to use a non-differentiable penalty such as the $\ell_1$-norm, 
i.e. the sum of the absolute values of the regression coefficients.  

\subsection{Our proposal: The mixture of self-regression with sparsity constraint}

In the present paper, we only intend to perform unsupervised clustering and thus, our setting seems 
far from the mixture of regression framework. The originality of the proposed approach is to 
introduce an artificial mixture of regressions for the simpler problem of clustering. The main 
idea is as follows: instead of estimating the $\mu_k's$ for $k=1,\ldots,K$, it should be easier to 
estimate only the coefficients of a sparse linear combinations of the $X_i's$ for all the datas 
belonging to the same cluster. This strategy should give even better results as the dimension $d$ 
of the problem increases if the sparsity of the involved linear combinations stays constant and small. 
Let us formalize our method in the next subsection. 

\subsubsection{The estimator}
In all what follows, we will assume that the data have been centered. 
Our proposal relies on the following simple idea: if the cluster proportions $p_k^*$'s, 
the class indices $Z_i$, $i=1,\ldots,n$ and the variances $\Sigma_k^*$, $k=1,\ldots,K$ where known ahead of time, 
the estimators of the $\mu_k$'s could be chosen, in the small sample setting, as linear combinations 
of the datas themselves. 

A simple example of such an idea is based on the notion of medoid. In clustering, the medoids play the role of 
the centers for each cluster, but are selected among the data themself. In what follows, instead of choosing 
only one medoid, we propose to select a linear combination of the data for each cluster. In order to stay 
robust as the dimension the space grows, we may impose that the linear combination be sparse, e.g. 
only onvolves 3 or 4 datas or less for each cluster. 
 
The main difficulty with such an approach is that choosing the right sample vectors to represent each cluster 
seems a priori a very hard task. Fortunately, one might rely on the recent discoveries concerning variable 
selection in order to overcome this problem: just as in the LASSO, a simple idea may be to use a 
regression formalism for estimating each $\mu_k$, $k=1,\ldots,K$, using a sparsity enforcing penalty like e.g. the 
$\ell_1$ norm of the coefficients. 

In the more general case where the indices $Z_i$, $i=1,\ldots,n$ are unobserved, and the cluster proportions $p_k^*$ and the covariance matrices $\Sigma_k^*$, $k=1,\ldots,K$ are unknown, one can 
consider maximizing the $l_1$-penalized log-likelihood like function given by 
\begin{equation}
\label{like}
\sum_{i=1}^n \log \Big(\sum_{k=1}^K p_k f^{(d)}(Y_i;\mu_k,\Sigma_k)\Big)-\lambda \sum_{k=1}^K \|\beta_k\|_1 
\end{equation}  
under the data-driven constraints $\mu_k=Y\beta_k$ for $k=1,\ldots,n$ where the matrix $Y$ is given by $Y=[Y_1,\ldots,Y_n]$. 
The parameter $\lambda$ is called the relaxation parameter. In other words, 
we would like to maximize the $l_1$-penalized likelihood function 
\begin{equation}
\tilde{l}_{pen}(\theta)=\tilde{l}(\theta)-\lambda \sum_{k=1}^K \|\beta_k\|_1,
\end{equation}
where 
\begin{equation}
\label{tildel}
\tilde{l}(\theta)=\sum_{i=1}^n \log \Big(\sum_{k=1}^K p_k f^{(d)}(Y_i;Y\beta_k,\Sigma_k)\Big).
\end{equation}

\subsubsection{The Space-Alternating $l_1$-EM algorithm}
\label{spal1}

Optimizing the $l_1$-penalized function (\ref{like}) can be performed using an EM-type algorithm. The Expectation Step consists of computing the conditional expectation 
of the complete $l_1$-penalized likelihood like function given the observations $Y_1,\ldots,Y_n$ where the distribution of the latent variables is taken to be their marginal density parametrized by the approximation $\bar{\theta}$ of the true parameter $\theta^*$. The resulting quantity is traditionally denoted by $Q(\theta,\bar{\theta})$ and we will use the same notation in our $l_1$-penalized context. 

More precisely, the complete $l_1$-penalized log-likelihood like function $\tilde{l}_{pen}^c(\theta)$, i.e. the penalized log-likelihood like function of the complete data $(Y_1,Z_1)$, \ldots, $(Y_n,Z_n)$ is given by 
\begin{equation}
\tilde{l}_{pen}^c(\theta) = \sum_{i=1}^n \log \Big(p_{Z_i} f^{(d)}(Y_i;Y\beta_k,\Sigma_k)  \Big)
-\lambda \sum_{k=1}^K\|\beta_k\|_1. 
\end{equation}
Thus, we obtain  
\begin{equation}
Q(\theta,\bar{\theta})=\sum_{i=1}^n \sum_{k=1}^K \log \Big(p_k f^{(d)}(Y_i;Y\beta_k,\Sigma_k)\Big) \tau_{i,k}
-\lambda \sum_{k=1}^K \|\beta_k\|_1
\end{equation}
where we used the standard notation $\tau_{i,k}=P_{\bar{\theta}}(Z_i=k\mid Y_1,\ldots,Y_n)$.  

The Maximization Step consists of maximizing $Q(\theta,\bar{\theta})$. 
In order to simplify the practical implementation, 
the $p_k$'s, $\beta_k$'s and $\Sigma_k$'s can be optimized alternatively in the manner of the Gauss-Seidel approach. 
In fact, the separability of the problem into two subproblems, the first being optimization over the $p_k$'s and the second being optimization over the $\beta_k$'s and $\Sigma_k$'s is already well known and the solution to the first of these 
subproblems is of the form 
\begin{equation}
p_k=\frac{\sum_{i=1}^n\tau_{i,k}}{\sum_{i=1}^n\sum_{k=1}^K \tau_{i,k}}.
\end{equation}
On the other hand, joint optimization in $\beta_k$'s and the $\Sigma_k$'s is not separable and space alternating option can be helpful in order to keep the computational complexity of each step at a low level. In order to address this problem, we need a generalization of the EM algorithm allowing for componentwise optimization at each step. Such penalized EM algorithms have been recently studied in the broader framework of Space Alternating Kullback 
Proximal Point Algorithms in \cite{Chretien:AISM08}. Optimizing successively over the $\beta_k$'s at one iteration and over the 
$\Sigma_k$'s at the next iteration should be reasonably efficient in most applications. 
Here, we will also optimize one cluster at a time in order to obtain the injectivity conditions which are needed
in the theoretical analysis of the algorithm. A simple way to accelerate the proposed version of the Gauss-Seidel 
methodology could be to average the new iterates $\beta^{(l)}$ and $\Sigma^{(l)}$ with the previous respective iterates so that to smooth the algorithm's trajectory. 

In what follows, we will restrict the analysis to the case where the covariance matrices are multiple of the 
identity but the method can easily be implemented and studied with general covariance matrices. 
The details of the method are summarized in Algorithm \ref{algo} below. The convergence analysis is provided in 
the Appendix (Section \ref{conv}).  
\vspace{.3cm}

\begin{algorithm}

\caption{\label{algo} Space-Alternating $l_1$-EM algorithm}
\begin{algorithmic}
\STATE {\bf Input} $L \in \mathbb N_*$         

\STATE Choose intial iterate $\theta^{(0)}=(p_1^{(0)},\ldots,p_k^{(0)},\beta_1^{(0)},\ldots,\beta_K^{(0)},\sigma_1^{(0)},\ldots,\sigma_K^{(0)})$  
\STATE $l=1$
\WHILE {$l\leq L$}
   \STATE (E-Step) Compute the conditional probabilities $P_{\theta^{(l-1)}}(Z_i=k\mid Y)$ given the observations $Y_1,\ldots,Y_n$ for $i=1,\ldots,n$ and $k=1,\ldots,K$ using the following formula
\begin{eqnarray}
\tau_{i,k}^{(l)} & = & \frac{
p_k^{(l-1)} f^{(d)} \left(Y_i;\mu_k^{(l-1)},\sigma_k^{(l-1)}I\right)  }
{\sum_{k=1}^K \ p_k^{(l-1)} f^{(d)} \left(Y_i;\mu_k^{(l-1)},\sigma_k^{(l-1)} I\right) }
\end{eqnarray}

{\bf compute}

\STATE --{\bf either} the $p_k^{(l)}$'s by the formula 
\begin{equation}
p_k^{(l)}=\frac{\sum_{i=1}^n\tau_{i,k}^{(l)}}{\sum_{i=1}^n\sum_{k=1}^K \ \tau_{i,k}^{(l)}}
\end{equation}

\STATE --{\bf or} $\beta_k^{(l)}$ as the solution of the LASSO-like optimization problem
\begin{equation}
\beta_k^{(l)} \in {\rm argmin}_{b\in \mathbb R^n} \|\left(\sum_{i=1}^n Y_i\tau_{i,k}^{(l)}\right)-Yb\|_2^2-\lambda \|b\|_1.
\end{equation}
for the index $k$ updated in cyclic order along iterations.  
\STATE --{\bf or} $\sigma_k^{(l)}$ using the formula 
\begin{equation}
\sigma_k^{(l)}=\frac1{\sum_{i=1}^k \tau_{i,k}^{(l)}} \sum_{i=1}^n \|Y_{i}-Y\beta_k^{(l-1)}\|_2^2 \: \tau_{i,k}^{(l)}.
\end{equation}
for one index $k$ updated in cyclic order along iterations.  

\STATE {\bf cyclically}

\ENDWHILE

\STATE {\bf Output} $p_k^{(L)}$, $\beta_k^{(L)}$ and $\sigma_k^{(L)}$ for $k=1,\ldots,K$.
\end{algorithmic}
\end{algorithm}

\section{Simulation results}
In this section, we address the question of testing the algorithm on simulated datasets.  
The Space Alternating $l_1$-EM was first tested on simulated data sets. The experiments were built as follows: 10 samples in $\mathbb R^2$ were generated from three 
different Gaussian distributions with the objective to recover the index of the distribution they were drawn from up to some index permutation. The class probabilities 
were taken as $p_1=.3$, $p_2=.2$ and $p_3=.5$ and the variances as $\sigma_1^2=5$, $\sigma_2^2=7$ and $\sigma_3^2=10$ without change through all the simulation 
experiments. Various experiments were performed using different values for the expectation vectors $\mu_1$, $\mu_2$ and $\mu_3$ since it could be easily suspected 
that the distance between them would play  a major role in the class index recovery problem. The results presented below were obtained 
using the following Monte Carlo scheme: the expectations were isotropic dilations of three points in $\mathbb R^2$ drawn uniformly at random in the cube $[-\frac12,\frac12]^3$. We 
ran the code for dilation factors $d$ going from 10 to 100 by steps of 10.

\subsection{Two dimensional data}
\begin{figure}[hbtp]
\begin{center}
\fbox{\begin{picture}(300.00,300.00)
\epsfig{file=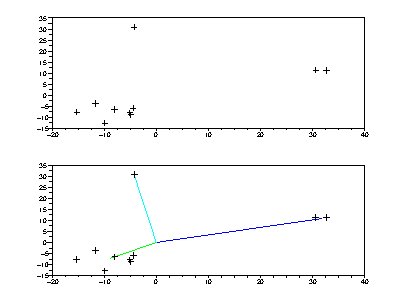,width=300.00pt,height=300.00pt}
\end{picture}}
\end{center}
\caption{\label{ex1} A result obtained with the LASSO (or space alternating $l_1$)-EM for centers drawn uniformly inside the cube $[-30,30]^3$.}
\end{figure}

An example of the type of result we obtained is given in Figure \ref{ex1} below where the 10 points were correctly classified.

Here is another example when the expectation vectors are chosen closer to each other and 8 points over 10 were correctly classified. 
\begin{figure}[hbtp]
\begin{center}
\fbox{\begin{picture}(300.00,300.00)
\epsfig{file=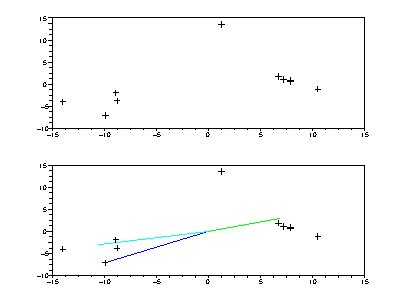,width=300.00pt,height=300.00pt}
\end{picture}}
\end{center}
\caption{\label{ex2} A result obtained with the space alternating $l_1$-EM for centers drawn uniformly in the cube $[-15,15]^3$.}
\end{figure}

The average number of correctly recovered class indices (ANCRCI) for 1000 Monte Carlo experiments in the case of 10 points as 
a function of the box into which the expectation vectors have been uniformly drawn are given in Table \ref{avnb} below.
\hspace{-2cm}
\begin{center}
\begin{table}[H]
\begin{tabular}{|c|c|c|c|c|c|}
\hline
Initial cube       & $[-5,5]^2$ & $[-10,10]^2$ & $[-15,15]^2$ & $[-20,20]^2$ & $[-25,25]^2$  \\
\hline 
ANCRCI             & 6.11 & 6.98 & 7.49 & 7.89 & 8.23  \\
\hline
Initial cube       & $[-30,30]^2$ & $[-35,35]^2$ & $[-40,40]^2$ & $[-45,45]^2$ & $[-50,50]^2$ \\
\hline
ANCRCI             & 8.41 & 8.38 & 8.61 & 8.71 & 8.74\\
\hline
\end{tabular}
\caption{\label{avnb} The average number of correctly recovered class indices (ANCRCI) over the 1000 Monte Carlo experiments 
similar to the one shown in Figure \ref{ex1} is given for increasing sizes 
of the initial cubes where the expectation vectors are chosen uniformy at random.}
\end{table}
\end{center}

\subsection{Higher dimensional data}
We performed Monte Carlo experiments in dimensions 5, 10 and 15. The results are 
presented in Table \ref{avnb}. In order to compare with the standard likelihood approach for finite Gaussian mixtures, 
we gathered the results obtained for the same experiments in Table \ref{gavnb} \footnote{We used the EM algorithm for Gaussian 
mixtures with covariance matrices equal to multiples of the identity matrix}. Table \ref{gavnbcem} shows the 
results obtained using the Classification EM (CEM) algorithm.

\begin{center}
\begin{table}[H]
\begin{tabular}{|c|c|c|c|c|c|c|c|}\hline
\hline
       & $[-20,20]^d$ & $[-25,25]^d$ & $[-30,30]^d$ &$[-35,35]^d$ & $[-40,40]^d$ & $[-45,45]^d$ & $[-50,50]^d$  \\
\hline 
$d=5$    &  8.18 & 8.37 & 8.54 & 8.6 & 8.69 & 8.73 & 8.63 \\
$d=10$   &  7.83 & 8.17 & 8.42 & 8.43 & 8.6 & 8.47 & 8.6 \\
$d=15$   &  7.58 & 8.07 & 8.15 & 8.25 & 8.35 & 8.34 & 8.42  \\
\hline
\end{tabular}
\caption{\label{avnb} The average number of correctly recovered class indices (ANCRCI) using our robust estimator over the 1000 Monte Carlo experiments shown in Figure{mc} is given for increasing sizes of the initial cubes where the expectation vectors are chosen uniformy at random and for increasing dimension of the sample space.}
\end{table}
%
\end{center}

\begin{center}
\begin{table}[H]
\begin{tabular}{|c|c|c|c|c|c|c|c|}\hline
\hline
      &  $[-20,20]^d$ & $[-25,25]^d$ & $[-30,30]^d$ &$[-35,35]^d$ & $[-40,40]^d$ & $[-45,45]^d$ & $[-50,50]^d$  \\
\hline 
$d=5$    &   6.75 & 6.86 & 6.9 & 7.06 & 7.11 & 7.04 & 7.07 \\
$d=10$   &   6.64 & 6.82 & 6.88 & 6.81 & 6.9 & 6.98 & 7.01 \\
$d=15$   &   6.56 & 6.65 & 6.93 & 6.85 & 6.8 & 6.9 & 7.01  \\
\hline
\end{tabular}
\caption{\label{gavnb} The average number of correctly recovered class indices (ANCRCI) using the standard maximum likelihood estimator (provided by the standard EM algorithm) over the 1000 Monte Carlo experiments is given for increasing sizes of the initial cubes where the expectation vectors are chosen uniformy at random and for increasing dimension of the sample space.}
\end{table}
\end{center}

\begin{center}
\begin{table}[H]
\begin{tabular}{|c|c|c|c|c|c|c|c|}\hline
\hline
      &  $[-20,20]^d$ & $[-25,25]^d$ & $[-30,30]^d$ &$[-35,35]^d$ & $[-40,40]^d$ & $[-45,45]^d$ & $[-50,50]^d$  \\
\hline 
$d=5$    &         8.055    &    8.27    &    8.36    &    8.37   &     8.33    &    8.42   &     8.41     \\ 
$d=10$   &          8.14    &    8.15    &    8.22    &    8.30    &    8.17   &     8.29   &     8.22  \\  
$d=15$   &          8.12    &    8.14    &    8.32    &    8.22    &    8.21   &     8.29   &     8.33     \\
\hline
\end{tabular}
\caption{\label{gavnbcem} The average number of correctly recovered class indices (ANCRCI) for the output of the Classification EM (CEM) algorithm, over the 1000 Monte Carlo experiments is given for increasing sizes of the initial cubes where the expectation vectors are chosen uniformy at random and for increasing dimension of the sample space.}
\end{table}
\end{center}

As Table \ref{avnb} shows, the class index recovery rate is still quite good in dimension 15 for well separed mixtures. 
A look at Table \ref{gavnb} shows that our method compares quite well with the standard likelihood approach for Gaussian mixtures estimation, 
especially in the higher dimensions where the average number of well classified data is better by often more than one unit.
The proposed $l_1$-penalized approach also compares favorably with the estimator given by  the CEM algorithm
as shown in Table \ref{gavnbcem}.  Giving a 
rigourous argument justifying these observations is currently under investigation but more experiments should be performed in order to explore in 
finer details the behavior of the method in more realistic context.

\section{Conclusion}
The goal of this paper was to propose a robust version of the maximum likelihood strategy for the estimation of finite Gaussian mixtures. Our approach is based 
on self-regression and sparse variable selection. Sparsity was promoted by using an $l_1$ penalty as in the LASSO. We developed a space alternating version of 
the penalized EM algorithm and proved that the interesting cluster points satisfy the Karush-Kuhn-Tucker optimality conditions. Our method was then 
tested on simulated datasets. In particular, the Monte Carlo experiments showed that cluster identification was more robust with our approach than by using the standard maximum likelihood estimator. Theoretical justifications of these observations ought to be investigated in a near future 
in order to increase our understanding of the strengths and weaknesses of this approach.  

{\bf Acknowledgement}. The author would like to thank Christophe Biernacki and Amelie Vaniscotte for very helpful discussions on the results of this paper. 

\section{Appendix: Convergence analysis of the Space-Alternating EM algorithm}
\label{conv}
In what follows, we will restrict our attention to the case where $\Sigma^2_k=\sigma^2 I$, $k=1,\ldots,K$ since 
more general forms of the covariance matrix will often be intractable in the small sample setting. 

When using a maximum likelihood approach, incorporation of a nondifferentiable penalty in the EM algorithm may cause some technical difficulties. 
A rigorous analysis has been proposed in \cite{Chretien:AISM08} in the case of general nondifferentiable penalties and space alternating 
optimization versions of the EM algorithm. The convergence analysis is made easier after interpreting the EM algorithm as a Proximal Point Algorithm 
which was first done in \cite{Chretien:IEEEIT00} (see also \cite{Chretien:ESAIMPS08} for more precise results). 

In our special case, we only need to show that our Space-Alternating $l_1$-EM is a 
Space-Alternating Kullback Proximal Point Algorithm of the form studied in \cite{Chretien:AISM08}
\footnote{for a definition of the Clarke subdifferential, see the Appendix of \cite{Chretien:AISM08}}.

\subsection{Recalls on Kullback-Proximal methods}

\subsubsection{Background}
Proximal point methods have been introduced by Martinet \cite{Martinet70} and Rockafellar \cite{Rockafellar76} 
in the seventies. A relationship between Proximal Point algorithms and EM algorithms 
was discovered in Chr\'etien and Hero (2000) (see also Chr\'etien and Hero (2008) for details). We review the EM analogy to KPP methods to motivate the space alternating generalization. 
Assume that a family of conditional densities $\{k(x| y;\theta)\}_{\theta \in {\mathbb R}^p}$ is such that the
Radon-Nikodym derivative $\frac{k(x| y,\bar{\theta})}{k(x| y;\theta)}$ exists
for all $\theta,\bar{\theta}$. We can define the following Kullback Leibler divergence:
\begin{equation}
\label{kullb}
I_y(\theta,\bar{\theta})={\sf E}\bigl[
\log \frac{k(x| y,\bar{\theta})}{k(x| y;\theta)}| y;\bar{\theta} \;
\bigr].
\end{equation}
Let $\Phi$ be a function to be maximized. Let us define $D_\phi$ as the domain of $\Phi$, $D_{I,\theta}$ the domain 
of $I_y(\cdot,\theta)$ and $D_I$ the domain of $I_y(\cdot,\cdot)$. Using the distance-like function $I_y$, the 
Kullback Proximal Point algorithm is defined by 
\begin{equation}
\label{proxit}
\theta^{k+1}={\rm argmax}_{\theta \in D_\Phi\cap D_{I,\theta}}\left\{\Phi(\theta)
-\beta_k I_y(\theta,\bar{\theta})\right\}.
\end{equation}
The following was proved in Chr\'etien and Hero (2000).
\begin{prop}{\rm [Chr\'etien and Hero (2000) Proposition 1].}
\label{equiem}
In the case where $\Phi$ is the log-likelihood, the EM algorithm is a special instance of the Kullback-proximal algorithm with 
$\beta_k =1$, for all $k\in \mathbb N$.
\end{prop}

\subsubsection{The Space Alternating Penalized Kullback-Proximal method}

In what follows, and in anticipation of component-wise implementations of penalized Kullback Proximal Point
algorithm, we will use the notation 
$\Theta_{\alpha}(\theta)$ for the local decomposition at $\theta$ defined by $\Theta_\alpha(\theta)=\Theta \cap \theta +\mathcal S_\alpha$, $\alpha =1,\ldots, R$ where $\mathcal S_1,\ldots,\mathcal S_R$ are subspaces of $\mathbb R^p$ and $\mathbb R^p=\oplus_{\alpha=1}^R \mathcal S_\alpha$. Let $\mathcal R$ be a list with finite cardinality $R$ and let $\rho$ denote any 
bijection from $\mathcal R$ to $\{1,\ldots,R\}$. 

Then, the Space Alternating Penalized Proximal Point Algorithm is defined as follows.
\begin{defi}
\label{SpAlt}
Let $l_y$ denote a function to be maximized (e.g. the log-likelihood or an arbitrary proxy). 
Let $\psi$: $\mathbb R^p \mapsto \mathcal S_1\times \cdots \times \mathcal S_R$ be a continuously differentiable mapping 
and let $\psi_{\alpha}$ denote its $\alpha^{th}$ coordinate. 
Let $(\nu_k)_{k\in \mathbb N}$ be a sequence of positive real numbers and $\zeta_r$, $r \in \mathcal R$ be non-negative real numbers.
Let $p_n$ be a nonnegative possibly nonsmooth locally Lipschitz penalty function with bounded Clarke-subdifferential on compact sets. Then, the Space Alternating Penalized Kullback Proximal Algorithm is defined by
\begin{equation}
\label{spakullprox}
\theta^{k+1}={\rm argmax}_{\theta\in \Theta_{k-1 ({\rm mod}\:\: R)+1}(\theta^k)\cap D_l\cap D_{I,\theta^k}}
\left\{l_y(\theta)-\sum_{r \in \mathcal R}\zeta_{r} p_n(\psi_{\rho(r)}(\theta))- \nu_k I_y(\theta,\theta^k)\right\},
\end{equation}
where $D_l$ is the domain of $l_y$ and $D_{I,\theta}$ is the domain of $I_y(\cdot,\theta)$.
\end{defi}

In most practical situations, the mappings $\psi_{\rho(r)}$ will simply be the projection onto the subspace $\Theta_{r}$, $r\in \mathcal R$.

\begin{prop}
The Space Alternating $l_1$-EM algorithm (defined in Section \ref{spal1}) 
is a particular instance of the Space Alternating Kullback Proximal Point Algorithm as defined by 
(\ref{spakullprox}).
\end{prop}

{\bf Proof}. 
First, we adopt the decomposition of the parameter space into the cartesian product of the $p_k$'s space, the $\beta_k$'s space and the $\sigma_k$'s 
space. More precisely $\Theta_1$ is the simplex in $\mathbb R^K$ and $\mathcal S_1=\mathbb R^K$, $\Theta_{2,k}=\mathbb R^n=\mathcal S_{2,k}$, and 
$\Theta_{3,k}=\mathbb R_+$ and $\mathcal S_{3,k}=\mathbb R$ for $k=1,\ldots,K$. Thus $r$ takes its values in the list 
$\mathcal R=\{1,(2,1),\ldots,(2,K),(3,1),\ldots,(3,K)\}$.

Then the mappings $\Psi_r$ are just the orthogonal projections onto $\mathcal S_r$ for $r \in \mathcal R$. 
Moreover $\zeta_1=0$ and $\zeta_{(3,k)}=0$ for $k=1,\ldots,K$ because the class probabilities and the variances are not penalized. Moreover we set $\zeta_{(2,k)}=\lambda$ for $k=1,\ldots,K$. 

Next, the $Q$-function can be written \footnote{see Section 4.1 of \cite{Chretien:AISM08} for more details.}
\begin{equation}
Q(\theta,\bar{\theta})=\tilde{l}(\theta)-I_y(\theta,\bar{\theta}) 
\end{equation} 
with \begin{equation}
\label{KLmix}
I_y(\theta,\bar{\theta})= \sum_{i=1}^n \sum_{k=1}^K t_{ik}(\bar{\theta})\log\Big(\frac{t_{ik}(\bar{\theta})}{t_{ik}(\theta)} \Big).
\end{equation}
where 
\begin{equation}
\label{toto}
t_{ik}(\theta)=\frac{p_k f^{(d)}(Y_i;Y\beta_k,\sigma_k I)}
{\sum_{l=1}^Kp_l f^{(d)}(Y_i;Y\beta_l,\sigma_l I)}.
\end{equation}
Thus, the space alternating LASSO-EM algorithm is a special case of the Space Alternating Kullback Proximal Point Algorithm 
for which the sequence $(\nu_k)_{k\in \mathbb N}$ is constant and the terms are all equal to one. 
\hfill$\Box$ 

We then have the following theorem. 
\begin{thm}
\label{bordp}
Let $\theta^*$ be a cluster point of the Space Alternating Penalized Kullback Proximal sequence. 
If $\theta^*$ lies in the interior of $D_{\tilde{l}}$, then $\theta^*$ satisfies the following property: there
exists a set of subsets $I_r^{**}\subset I^*$ where $I^*$ denotes the index of the active constraints at $\theta^*$, i.e. 
$I^*=\{(i,j) \textrm{ s.t. } t_{i,j}(\theta^*)=0$, and there is a family of real numbers $\gamma_{ij}$, 
$(i,j)\in \mathcal I_r^{**}$ , $r \in \mathcal R$ such that the following Karush-Kuhn-Tucker condition for optimality holds at cluster point $\theta^*$: 
$$
0\in  \nabla \tilde{l}(\theta^*)-\sum_{r \in \mathcal R}\zeta_{r} \partial p_n(\psi_{\rho(r)}(\theta^*))
+\sum_{r\in \mathcal R} \sum_{(i,j)\in \mathcal I_r^{**}} \gamma_{ij}^*\nabla t_{ij}(\theta^*).
$$
\end{thm}
{\bf Proof}. 
We start by verifying that Assumptions 2.2.1, 2.2.3 and Assumptions 2.2.4 of \cite{Chretien:AISM08} hold in our case. 
The differentiability requirement in Assumptions 2.2.1.(i). is obvious. However, if one $\beta_k$ belongs to the kernel of $Y$, it may be of any arbitrary large norm without leading the log-likelihood towards $-\infty$. However, note that, as is well 
known in Gaussian mixture models, $\tilde{l}$ tends to $+\infty$ only at finite number of degenerate points. Thus, since, the penalization terms $p_n$ tend to $+\infty$ as the norm of any $\beta_k$ tends to $+\infty$, the 
function $Q(\theta,\theta^{(k)})$ tends to $-\infty$ if the norm of any $\beta_k$ goes to $+\infty$. Moreover, 
as easily checked on the expression of the likelihood, the function $Q(\theta,\theta^{(k)})$ also 
goes to $-\infty$ if any variance $\sigma_k^2$ goes to $+\infty$. 

The domain $D_{\tilde{l}}$ is defined by the fact that the term inside the log in (\ref{lkhd})
must be positive. On the other hand, for any $\bar{\theta}$ in $\Theta=\Theta_1\times \Theta_{2,1}\times \cdots \times \Theta_{2,K}\times \Theta_{3,1}\times \cdots \times \Theta_{3,K}$,
the domain $D_{I_y,\bar{\theta}}$ is the set of the $\theta$'s for which the $t_{ik}(\theta)$ are positive, and therefore, 
does not depend on $\bar{\theta}$. Moreover, the set of $\theta$'s for which the $t_{ik}(\theta)$ are positive is $D_{\tilde{l}}$. 
Thus, the projection of $D_{I}$ onto the first coordinate is $D_{\tilde{l}}$ and Assumptions 2.2.1.(ii). are satisfied. 

Assumptions 2.2.1.(iii). is immediate since here the relaxation sequence (denoted here by $(\nu_k)_{k\in\mathbb N}$) is constant. Assumptions 2.2.1.(iv). is also straightforward since the mappings $\Psi_r$ are orthogonal projections onto $\mathcal S_r$, 
$r\in \mathcal R$. 

In our context, based on (\ref{toto}), we have $\phi=t\log(t)-1$ and Assumptions 2.2.3.(i)-(iii). are easily verified.  Injectivity 
of the mapping $t$ when restricted to $\cup_{j=1}^3 \Theta_{j,k}$ is proved in \cite{Celeux:JCGS01} and thus, injectivity holds on each $\Theta_{1,k}$,\ldots,$\Theta_{3,k}$ and Assumption 2.2.3.(iv) holds. 

Moreover, since $t_{ik}(\theta)=0$ implies that $p_k=0$
and $p_k=0$ implies
\begin{equation}
\frac{\partial t_{ik}}{\partial \beta_{jl}}(\theta)=0
\end{equation}
for all $j=1,\ldots,p$ and $l=1,\ldots,K$ and 
\begin{equation}
\frac{\partial t_{ik}}{\partial \sigma^2}(\theta)=0,
\end{equation}
it follows that $P_{{\mathcal S}_r}(\nabla t_{ik}(\theta^*))=\nabla t_{ik}(\theta^*)$ if $\mathcal S_r$ is the vector space generated by 
the probability vectors $p$ and $P_{{\mathcal S}_r}(\nabla t_{ik}(\theta^*))=0$ otherwise. 

Let $\theta^*$ be a cluster point in the interior of $D_{\tilde{l}}$. Since the $t_{ik}$ are 
clearly continuously differentiable around such a $\theta^*$, Corollary 1 in \cite{Chretien:AISM08}
gives that $\theta^*$ satisfies the following property: there
exists a set of subsets $I_r^*\subset I^*$ and a family of real numbers $\gamma_{ij}$, 
$(i,j)\in \mathcal I_r^*$ , $r \in \mathcal R$ such that the following Karush-Kuhn-Tucker condition for optimality holds at cluster point $\theta^*$: 
$$
0\in  \nabla \tilde{l}(\theta^*)-\sum_{r \in \mathcal R}\zeta_r \partial p_n(\psi_r(\theta^*))
+\sum_{r\in \mathcal R} \sum_{(i,j)\in \mathcal I_r^{**}} \gamma_{ij}^*\nabla t_{ij}(\theta^*),
$$
which is the desired result.
\hfill$\Box$

The meaning of this theorem is simply that a Karush-Kuhn-Tucker condition is satisfied at any cluster point in the domain of definition of the 
log-likelihood.




\begin{thebibliography}{00}



\bibitem{Biernacki:CSDA06} C. Biernacki, G. Celeux, G. Govaert and F. Langrognet, (2006) "Model-based cluster and discriminant analysis with the MIXMOD software", Computational Statistics and Data Analysis, Vol. 51, 2, 587--600.

\bibitem{Biernacki:SPL03} C. Biernacki and S. Chr\'etien, (2003) "Degeneracy in the maximum likelihood estimation of univariate Gaussian mixtures with EM", Statist. Probab. Lett. 61, no. 4, 373--382.

\bibitem{Candes:AS09} E. Cand\`es and Y. Plan, (2009) "Near ideal model selection by $l_1$ penalization", The Annals of Statistics, to appear. 

\bibitem{Celeux:JCGS01} G. Celeux, S. Chr\'etien, F. Forbes and A. Mkhadri (2001) ``A Component-Wise EM Algorithm for Mixtures''
Journal of Computational and Graphical Statistics, vol. 10, no. 4, 697-712. 

\bibitem{Chretien:IEEEIT00} S. Chr\'etien and A. Hero,  (2000) ``Kullback proximal algorithms for maximum-likelihood estimation''. Information-theoretic imaging.  IEEE Trans. Inform. Theory  46,  no. 5, 1800--1810

\bibitem{Chretien:ESAIMPS08} S. Chr\'etien and A. Hero, (2008) `` On EM algorithms and their proximal generalizations''. ESAIM P\&S  12, 308--326

\bibitem{Chretien:AISM08} S. Chr\'etien, A. Hero and H. Perdry (2008) ``Space Alternating Penalized Kullback Proximal Point Algorithms for Maximing Likelihood with Nondifferentiable Penalty'', Annals Inst. Stat. Math., to appear. 
Available at http: //arxiv.org/abs/0901.0017


\bibitem{Dempster:JRSS77} A.~P. Dempster, N.~M. Laird, and D.~B. Rubin, (1977) ``Maximum likelihood from
  incomplete data via the {EM} algorithm,'' {\em J. Royal Statistical Society,
  Ser. B}, vol. 39, no. 1, pp. 1--38.

\bibitem{Fan:JASA01} J. Fan and R. Li (2001) ``Variable selection via non-concave penalized likelihood and its oracle properties'', Journal of the American 
Statistical Association, 96, 1348--1360.

\bibitem{Fessler:IEEESigProc94}  J.~A. Fessler, and  A.~O. Hero, (1994)``Space-alternating generalized expectation-maximization algorithm'', IEEE Trans. Signal Processing,
vol. 42,  no. 10, pp.  2664--2677.


\bibitem{Hurn:JCGS03} M. Hurn M, A. Justel and C.P. Robert, (2003) "Estimating Mixtures of Regressions", Journal of Computational and Graphical Statistics, 12, 55--79.



\bibitem{Khalili:JASA07} A. Khalili and J. Chen, (2007) ``Variable Selection in Finite Mixture of Regression Models'', Journal of the American Statistical Association, Volume 102, Number 479, pp. 1025-1038.



\bibitem{McLachlan:Book00} G.J.~McLachlan and D. Peel. (2000) {\em Finite Mixture Models}. Wiley

\bibitem{Martinet70}  B.~Martinet (1970). R\'egularisation d'in\'equation variationnelles par
  approximations successives. {\em Revue Francaise d'Informatique et de
  Recherche Operationnelle}, vol. 3, pp. 154--179.

\bibitem{Rockafellar76}  R.~T. Rockafellar (1976). Monotone operators and the proximal point algorithm. {\em SIAM Journal on Control and Optimization}, vol. 14, pp. 877--898.

\bibitem{Tibshirani:JRSS96} R. Tibshirani, (1996) `` Regression shrinkage and selection via the LASSO'', Journal of the Royal Statistical Society, Series B, vol. 58, no. 1, 
pp. 267--288. 



\end{thebibliography}
\end{document}